# Observation of novel robust edge states without bulk-boundary correspondence in non-Hermitian quantum walks

Bo Wang,1* Tian Chen,1*,+ and Xiangdong Zhang 1,$

*1Beijing Key Laboratory of Nanophotonics & Ultrafine Optoelectronic Systems, School of Physics, Beijing Institute of Technology, 100081, Beijing, China*
*These authors contributed equally to this work.
$Corresponding author: zhangxd@bit.edu.cn
+Corresponding author: chentian@bit.edu.cn

## Abstract

Recently, the study of non-Hermitian physics has attracted considerable attention. The modified bulk-boundary correspondence has been proposed to understand topological edge states in non-Hermitian static systems. Here we report a new experimental observation of edge states in non-Hermitian periodically driven systems. Some unconventional edge states are found not to be satisfied with the bulk-boundary correspondence when the system belongs to the broken parity-time (PT) symmetric phase. The experiments are performed in our constructed non-Hermitian light quantum walk platform with left and right boundaries, where the beams outside system boundary are blocked subtly at the end of each step. The robust properties of these edge states against to static perturbations and disorder have also been demonstrated experimentally. The finding of robust edge states in broken PT-symmetric phase inspires us to explore a robust transport channel in ubiquitously complex systems with strong dissipation.

# 1. Introduction

The emergence of robust edge states plays an important role in topological physics, since it has a potential application in designing robust unidirectional transport devices and realizing topological quantum computation without decoherence [1]. In static Hermitian systems [2-4], the robust edge states at the interface between two systems with different topological phases can be predicted correctly by the bulk topological invariants in the band-theory framework, which is called as the bulk-boundary correspondence [5, 6]. The bulk-boundary correspondence can be applied universally to all noninteracting static Hermitian lattices.

In many real systems, however, physical properties should be described by non-Hermitian Hamiltonians due to dissipation. Recently, the studies on topological edge states have been extended to non-Hermitian systems. Many interesting topological phenomena associated with edge states have been revealed, e.g., skin effect, lasing and so on [7-21], which cannot be observed in Hermitian systems. In these cases, some revised topological invariants are presented to characterize the topological phases, that is, the modified bulk-boundary correspondence has been proposed to understand the phenomena [14-21]. The remarkable feature of non-Hermitian systems is that there is a unique point in parameter space, the so-called exceptional point (EP) [22, 23], where the underlying modes coalesce to a single mode. On one side of EP in the parameter space, the system belongs to the unbroken parity-time (PT) symmetric phase with all eigenmodes being entirely real; on the other side of EP, the system is in the broken PT-symmetric phase accompanied by strong dissipation in the eigenmodes. The bi-orthonormal basis for the non-Hermitian system is introduced to analyze new phenomena [22]. The previous studies [7-13] have mainly focused on robust edge states at the interface between two topologically distinct unbroken PT-symmetric phases. The question about the existence of robust edge states in the boundaries of system with only broken PT-symmetric phase remains open.

In this work, we report experimental observations of edge states in dissipative non-Hermitian systems with left and right boundaries. The experiments are performed using our constructed non-Hermitian quantum walk (QW) platform. Recent investigations [24-35] have shown that QWs can provide a versatile platform for studying the topological phase of time-driven system. Based on such platforms, the topologically protected edge states at the interface between regions with different bulk topological phases have been observed not only in Hermitian systems [28-33], but also in the non-Hermitian systems with unbroken PT-symmetric phase [36-38]. These robust edge states have been proved to be described by conventional topological invariants [37, 38]. However, our experimental results show that the number of edge states has a drastic change when the system changes from the unbroken PT-symmetric phase to the broken PT-symmetric phase. The increase or decrease of edge state number does not agree with the corresponding change of topological invariant for the bulk, which means that the bulk-boundary correspondence cannot work. Our experimental implementation of the non-Hermitian QW contributes towards a new way to study the novel phenomena of Floquet topological insulator [39] and opens up the exciting possibility of exploring a robust transport channel in complex systems with strong dissipation.

## 2. Experimental realization of non-Hermitian QWs with left and right boundaries.

Our experimental setup to perform QWs is illustrated in Fig. 1, which consists of three modules: Fig. 1(a), state preparation; Fig. 1(b1) and 1(b2), many steps of QWs with left and right boundaries, respectively; Fig. 1(c), detection. In our experiments, the classical light source (from a 632.8 nm helium–neon laser) is used, since the previous investigations have shown that QWs can be implemented using an entirely classical light source due to the similarity between coherent processes in quantum mechanics and classical optics [40, 41]. A coin and one-dimensional positions of QWs are encoded by the polarization and spatial positions of beam, respectively.

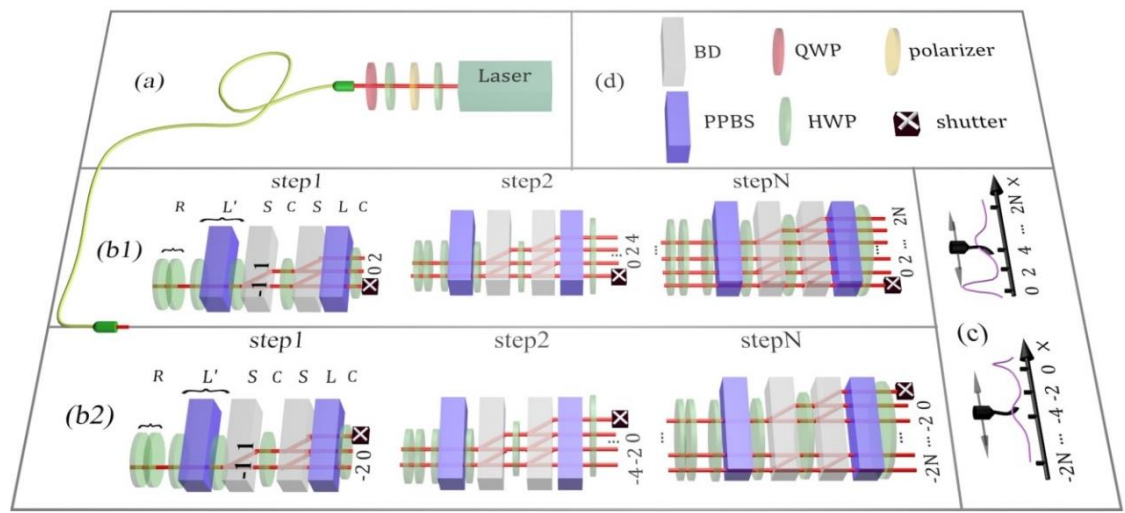

**Fig. 1**. (Color online) Experimental scheme for non-Hermitian QWs with left and right boundaries. (a) State preparation. (b) Many steps of QWs with boundaries. To study novel edge states in non-Hermitian QWs with boundaries, we create left and right boundaries in (b1) and (b2), respectively. (c) Detection. (d) Optical elements.

The beam with arbitrary initial polarization and intensity can be obtained in Fig 1(a), which will enter the module of many steps of QWs. In order to study transport behaviors at edges, we construct the non-Hermitian QWs with left (Fig. 1(b1)) or right (Fig. 1(b2)) boundary, where the beam is guided to a series of optical devices which consist of many step operators. One step operator of QW is described by $U_L$ with left or right boundary. The operator $U_L$ is expressed as

$$U_L = C(\theta_1/2) L S C(\theta_2) S L' R(\theta_1/2), \tag{1}$$

where $C(\theta) = \sum_x |x\rangle\langle x| \otimes e^{-i\theta\sigma_y} \sigma_z$ is a coin operator, which is realized by one HWP with rotation angle $\theta$ on the polarization state. Another coin operator $R(\theta) = C(\theta)C(0)$ is realized by a combination of two HWPs. Here $\sigma_y$ and $\sigma_z$ are Pauli operators in the $y$ and $z$ directions, respectively. In Eq. (1), $S = \sum_x |x+1\rangle\langle x| \otimes |H\rangle\langle H| + |x-1\rangle\langle x| \otimes |V\rangle\langle V|$ represents the conditional shift operator along the $x$-direction, which is realized by beam displacers (BDs). $|H\rangle = (1,0)^T$ and $|V\rangle = (0,1)^T$ stand for the horizontal and vertical polarization state, respectively. In the experiment, we set the initial position state for the walker as $|0\rangle$. Different output spatial positions on the lateral sections of BDs are used to represent different sites in the QWs. The loss operators $L$ and $L'$ have the form as

$$L = \sum_x |x\rangle\langle x| \otimes \begin{pmatrix} l_1 & 0 \\ 0 & l_2 \end{pmatrix}, \quad L' = \sum_x |x\rangle\langle x| \otimes \begin{pmatrix} l_2 & 0 \\ 0 & l_1 \end{pmatrix}, \tag{2}$$

where $0 \leq l_1, l_2 \leq 1$ represent different losses for the beams with different polarizations. When corrected by one factor, the QW $U_L$ changes to

$$U = U_L / l_1 l_2 = C\left(\theta_1/2\right) G S C(\theta_2) S G^{-1} R\left(\theta_1/2\right). \tag{3}$$

The gain-loss operator $G$ is given by $G = \sum_x |x\rangle\langle x| \otimes \begin{pmatrix} e^r & 0 \\ 0 & e^{-r} \end{pmatrix}$, and the gain-loss strength is $r = \frac{1}{2}\ln(l_1/l_2)$. In this way, the effective "gain" and "loss" emerge at the walker with different polarizations. Experimental implementation of alternating gain-loss with partially polarizing beam splitter (PPBS) in the QWs is shown in S1 of Supplementary Materials. As shown in Fig. 1(b1), the QW system with the left boundary is obtained as follows: when the beam is emitted from the last HWP of each step, we use a shutter to block part of beam with its position state $|x\rangle(x<0)$. With this blockade operation, the element $U_{(m,t),(n,t')}(m<n, n=0; t,t'=|H\rangle \text{ or } |V\rangle)$ describing the transition from $|n=0\rangle|t'\rangle$ to $|m\rangle|t\rangle$ in the evolution operator $U$ is equal to zero, where the subscripts $m$ and $n$ represent the indices of positions in the $x$-direction. It means that the walker standing at the left boundary $(x=0)$ cannot reach the position $x<0$ at the end of each step. Therefore, the evolution into the position $x<0$ is prohibited and the QW with the left boundary is constructed. Similar to this construction, in Fig. 1(b2) we block the beam representing the position states of walker $|x\rangle(x>0)$ at the end of each step and one non-Hermitian QW with the right boundary is obtained. The detail constructions of the non-Hermitian QWs with left and right boundaries are given in S1 of Supplementary Materials. After undergoing N steps of QW, the beams transport into the detection module in Fig. 1(c). Herein, the output light intensity in each step can be measured by the power meter. The list of optical devices is shown in Fig. 1(d).

**3. Experimental observation of edge states in non-Hermitian QWs.**

In fact, the discussions on edge states in the non-Hermitian QWs have been done in Ref. [37], which only focused on bound states at the interface between two unbroken PT-symmetric phases. It is shown that the appearance of the bound states can be dictated by the bulk-boundary correspondence. Here we study topological properties of non-Hermitian QW systems surrounded with the vacuum using the platform shown in Fig. 1. Based on this platform, we want to study the behavior of edge states and explore whether the bulk-boundary correspondence exists when the system changes from unbroken PT-symmetric phase to broken PT-symmetric phase. We first calculate the eigenvalues $\lambda = e^{i\varepsilon}$ of $U$ with left and right boundaries for such non-Hermitian QWs as a function of the gain-loss strength $r$ with fixed $(\theta_1, \theta_2)$. We set the non-Hermitian QW $U$ with left and right boundaries simultaneously. Therefore, we can obtain the corresponding edge states at left and right boundaries simultaneously.

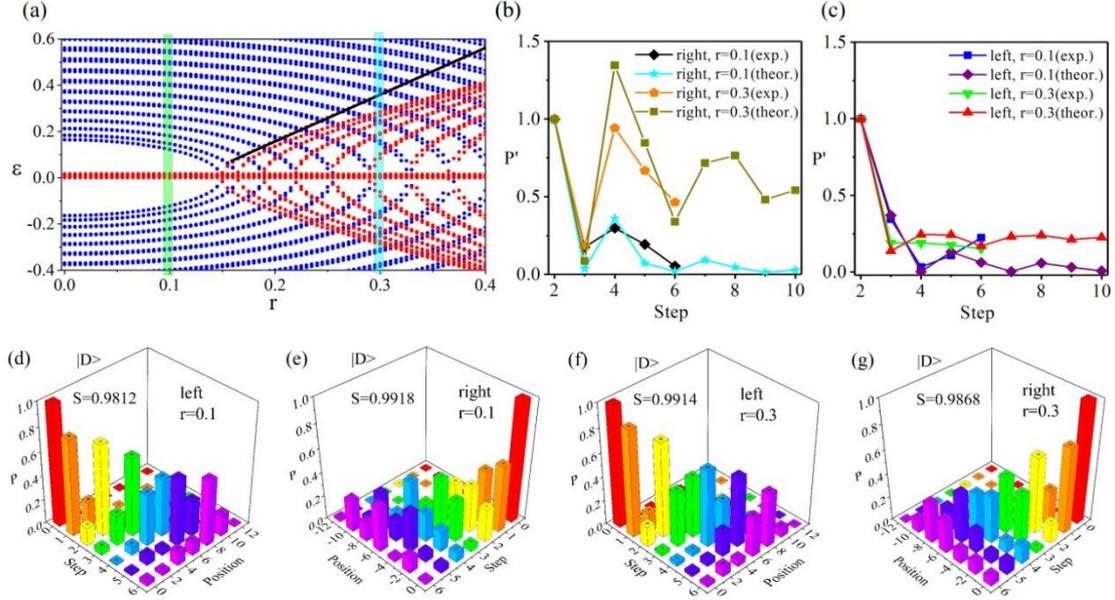

**Fig. 2.** Experimental observation of edge states in non-Hermitian QWs with six steps at $\theta_1 = 0.16\pi$ and $\theta_2 = 0.21\pi$. (a) The energy spectrum $\lambda = e^{i\varepsilon}$ for the non-Hermitian QW. The real and imaginary parts of quasienergies $\varepsilon$ are denoted by blue and red dots. Isolated degenerated eigenvalues are emphasized with black solid lines. The gain-loss strengths $r = 0.1$ in green region and $r = 0.3$ in cyan region, respectively. The initial polarization of walker is $|D\rangle = \frac{1}{\sqrt{2}}(|H\rangle + |V\rangle)$. The walker starts from the left (right) boundary of system in (d) and (f) [(e) and (g)]. In (b) and (c), the probabilities at the right and left boundary normalized to the localized probability at the second step $P' = P/P(step = 2)$ are provided. Black rhombus (cyan stars) and orange pentagons (dark yellows squares) in (b) stand for the experimental (theoretical) localized probabilities at right boundary for (e) and (g), respectively. Blue squares (purple rhombus) and green down-triangles (red up-triangles) in (c) stand for the experimental (theoretical) localized probabilities at right boundary for (d) and (f), respectively.

In Fig. 2(a), we provide the calculated results for quasienergy $\varepsilon$ at the rotation angles $\theta_1 = 0.16\pi$ and $\theta_2 = 0.21\pi$. Since the non-Hermitian QWs satisfy particle-hole symmetry, the energy spectrum also shows such a symmetry. The related analysis is shown in S2 of Supplementary Materials. It is seen clearly from Fig. 2(a) that the EP point appears at $r = 0.15$. When $r < 0.15$, the system belongs to the unbroken PT-symmetric phase and no isolated eigenvalue is found. However, when $r > 0.15$, the system enters the broken PT-symmetric phase and two isolated degenerated eigenvalues appear (black line). These isolated degenerated eigenvalues correspond to two edge states which locate at the right boundary of system. Details of distribution of eigenstates are given in S3 of Supplementary Materials.

In order to test theoretical results, we perform the corresponding QW experiment according to the scheme shown in Fig. 1. First, we adjust the rotation angles $(\theta_1, \theta_2)$ to the corresponding case. Then, by modulating transmissivity parameters in the PPBS, we can easily change the system from the unbroken PT-symmetric phase to broken PT-symmetric phase. For example, the parameters $(l_1^2, l_2^2)$ change from $(1, 0.67)$ to $(1, 0.30)$, which correspond to the gain-loss strength $r = 0.1$ and 0.3, respectively.

In Fig. 2(d) and 2(e), we show the experimental results of probability distributions of non-Hermitian QWs at $r = 0.1$. The corresponding results at $r = 0.3$ are given in Fig. 2(f) and 2(g). The similarity $S[P_{th}(x), P_{ex}(x)] = \left( \sum_{x=-N}^{N} \sqrt{P_{th}(x) P_{ex}(x)} \right)^2$ is presented to evaluate the quality of experiment. Here, $P_{th}(x)$ ($P_{ex}(x)$) represents the theoretical (experimental) probability distribution in QWs. The values of $S$ obtained in Fig 2(d)-2(g) are larger than 0.98, which indicates that the experiment results agree well with the theoretical results. The analysis of error has been presented in S1 of Supplementary Material. It is found that no matter whether the walker starts from left (Fig. 2(d)) or right (Fig. 2(e)) boundary, the probability distributions at $r = 0.1$ always decrease quickly with increase of steps, and most of probability distributions are in the center of position space of system. In contrast, the situation becomes different at $r = 0.3$. In this case, the amplitude of probability located at the right boundary keeps a relative large value in the last few steps (Fig. 2(g)), while the probability at the left boundary decreases quickly (Fig. 2(f)). This can be seen more clearly from Fig. 2(b) and 2(c).

In Fig. 2(b) and 2(c), we plot localized probabilities with steps at the boundary $x = 0$. Figure 2(b) displays experimental and theoretical results at the right boundary corresponding to Fig.

2(e) and 2(g), and Fig. 2(c) shows the case at the left boundary (Fig. 2(d) and 2(f)). It is seen clearly that the localized probabilities at the right boundary with $r=0.3$ are nearly one order of magnitude larger than those with $r=0.1$ when the step of QW $n \geq 4$, which indicates that there exist edge states at the right boundary with $r=0.3$. However, the probabilities located at the left boundary reduce quickly to a very small value ($P'<0.2$) for both $r=0.1$ and $r=0.3$. This means that no edge state exists at these cases.

The above results only focus on a sort of fixed rotation angles. In fact, as the rotation angle of QW changes, so does the phenomenon. In Fig. 3(a), we plot the energy spectrum at $\theta_1 = 0.3\pi$ and $\theta_2 = 0.16\pi$. Details of distribution of eigenstates can be found in S3 of Supplementary Materials. The EP point appears at $r=0.34$. When the parameters r alter from 0.3 to 0.4, the number of the isolated eigenvalue changes from 4 to 2. The parameters $(l_1^2, l_2^2)$ change from $(1, 0.30)$ to $(1, 0.20)$, respectively. The system also changes from the unbroken PT-symmetric phase to the broken PT-symmetric phase as similar as Fig 2(a).

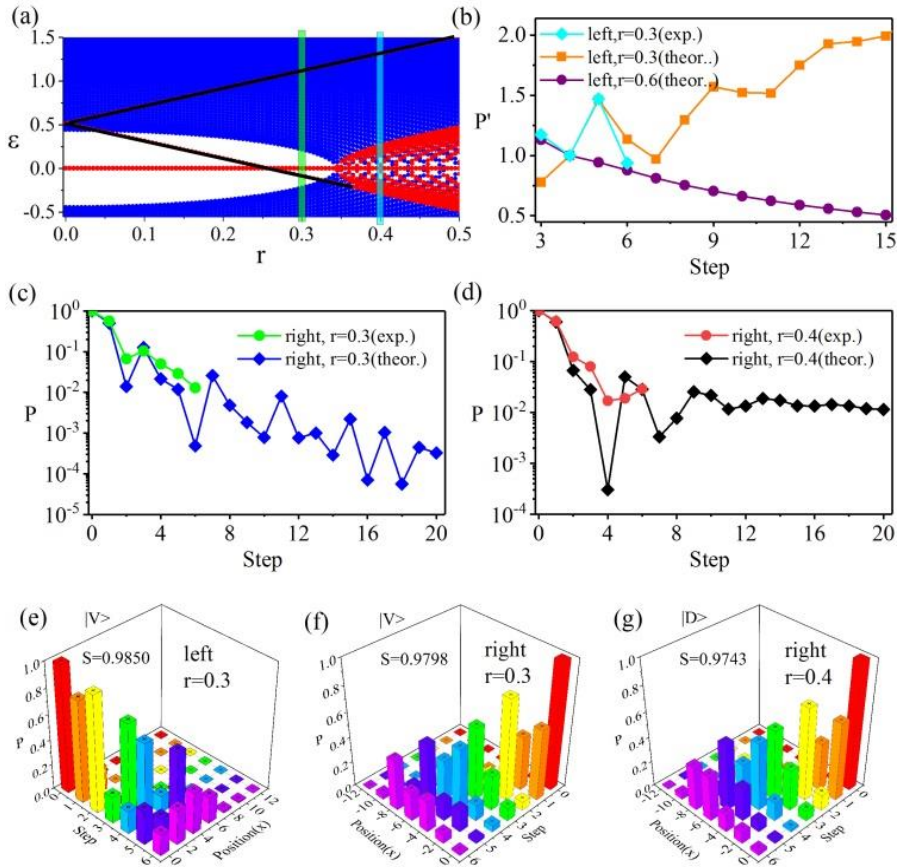

**Fig. 3**. Experimental observation of edge states in non-Hermitian QWs with six steps at $\theta_1 = 0.3\pi$ and $\theta_2 = 0.16\pi$. (a) The energy spectrum $\lambda = e^{i\varepsilon}$ for the non-Hermitian QW. The real and imaginary part of quasienergies $\varepsilon$ are denoted by blue and red dots. Isolated degenerated eigenvalues are emphasized with black solid lines. The gain-loss strengths $r = 0.3$ in green region and $r = 0.4$ in cyan region. The walker starts from left (right) boundary of system in (e) [(f) and (g)]. The initial polarization of walker: $|V\rangle$ for (e) and (f); $|D\rangle$ for (g), respectively. The gain-loss strengths $r = 0.3$ in (e) and (f); $r = 0.4$ in (g). In (b), the probabilities at the left boundary normalized to the localized probability at the fourth step $P' = P/P(step = 4)$ are provided. Cyan rhombus stands for the experimental results with $r = 0.3$. Brown rectangles and purple circles represent the theoretical results with $r = 0.3$ and $r = 0.6$. In (c) and (d), the probabilities at the right boundary at $r = 0.3$ and $r = 0.4$ are provided. Green (red) circles and blue (black) rhombus represent the experimental and theoretical results with $r = 0.3$ ($r = 0.4$).

Figure 3(e) and 3(f) show the experimental results of probability distributions of non-Hermitian QWs at $r = 0.3$. When the walker starts from the left boundary (Fig. 3(e)), the amplitude of probability located at the boundary increase to a large value step by step. In order to clearly show this phenomenon, in Fig. 3(b) we plot localized probabilities at the boundary $x = 0$. The amplification of probability at the boundary means that edge states enhance the localization with steps. This enhancement can be attributed to the corresponding eigenvalues $|\lambda|$ (in this case, the imaginary part of $\varepsilon$ smaller than 0) for these edge states larger than 1. For comparison, we also study the theoretical corrected probabilities at the left boundary with $r = 0.6$. In such a case, the system belongs to the broken PT-symmetric phase and the original two edge states at the left boundary disappear (the lower black line enters the red region with $r = 0.6$ in Fig. 3(a)). The non-existence of these edge states makes the localized probability at the left boundary decreases only (purple circles in Fig. 3(b)).

In contrast, the located probability at the right boundary of $r = 0.3$ shows a downward trend with increase of step (Fig. 3(f)). This is similar to the case at $r = 0.4$ (Fig. 3(g)). The corresponding probabilities for $r = 0.3$ and $r = 0.4$ at $x = 0$ are given in Fig. 3(c) and 3(d), respectively. In these two cases, edge states at the right boundary have the corresponding eigenvalue $|\lambda|$ (in this case, the imaginary part of $\varepsilon$ larger than 0) smaller than 1. Thus, the localizations at the right boundary induced by these edge states are weakened during the evolution. Although these localized probabilities display the downward trend, they oscillate around a non-zero value during the evolution, which correspond to the edge states at the right boundary (upper black line in Fig. 3(a)). Furthermore, we also test the robust properties of these edge states. The experimental results show that these edge states are robust against small perturbations and disorder. The detailed discussions and results are given in S4 of Supplementary Materials.

**4. Characterizing robust edge states with current theories.**

In order to describe physical properties of robust edge states in non-Hermitian systems, some typical theories on topological invariants have been developed in recent years, such as the complex phase for energy bands [17], the vorticity of energy eigenvalues [18], "Q-Matrix" [19], the vorticity of topological defects [20], and the step-average of mean position of walker [38]. In the following, we present the discussion with the complex phase. By employing the Fourier transform, the non-Hermitian QW $U$ can be expressed in its momentum space as

$$U = \int \frac{dk}{2\pi} U_k \otimes |k\rangle\langle k|, U_k = d_0 \sigma_0 + d_x \sigma_x + i d_y \sigma_y + i d_z \sigma_z,$$

with $d_0 = \cos 2k \cos\theta_1 \cos\theta_2 - \cosh 2r \sin\theta_1 \sin\theta_2$, $d_x = -\sinh 2r \sin\theta_2$,

$d_y = -\cos\theta_2 \sin\theta_1 \cos 2k - \cosh 2r \cos\theta_1 \sin\theta_2$ and $d_z = \cos\theta_2 \sin 2k$. To simplify the calculation, we apply the unitary operation to the evolution operator $U_k$, that is $\tilde{U}_k = \exp\left(-i\frac{\pi}{4}\sigma_y\right) U_k \exp\left(i\frac{\pi}{4}\sigma_y\right)$. We obtain the left and right eigenstates [22, 37] for the system as $\langle\varphi_\pm^L| = \frac{1}{\sqrt{2\cos 2\Omega_k}}\left(e^{\pm i\Omega_k} \quad \mp i e^{i\theta_k} e^{\mp i\Omega_k}\right)$ and $|\varphi_\pm^R\rangle = \frac{1}{\sqrt{2\cos 2\Omega_k}}\begin{pmatrix} e^{\pm i\Omega_k} \\ \pm i e^{-i\theta_k} e^{\mp i\Omega_k} \end{pmatrix}$, with

$d_y + i d_z = |d(k)| e^{i\theta_k}$ and $d_x = |d(k)| \sin 2\Omega_k$. Details of calculation results have been presented in S5 of Supplementary Materials. The subscripts + and − represent the energy bands with eigenvalues $E_{k,+}$ and $E_{k,-}$ of $\tilde{U}_k$ ($U_k$), respectively. Then we define the complex phase as [17]

$$v_{tot} = \frac{1}{2}(v_+ + v_-) = -\frac{1}{2\pi}\int d\theta_k. \tag{4}$$

With $v_\pm = \frac{-i}{\pi}\int dk \langle \varphi_\pm^L | \frac{d}{dk} | \varphi_\pm^R \rangle$. The calculated results for the complex phase with $r=0.3$ and $r=0.1$ are shown in Fig.4 (a) and (b) in non-Hermitian QWs with rotation angles $\theta_1$ and $\theta_2$. The black and orange regions represent the complex phase values 0 and -2, respectively. The corresponding results for the number of edge states are shown in Fig. 4(c) and (d). The black, brown and orange regions denote the value 0, 2 and 4, respectively.

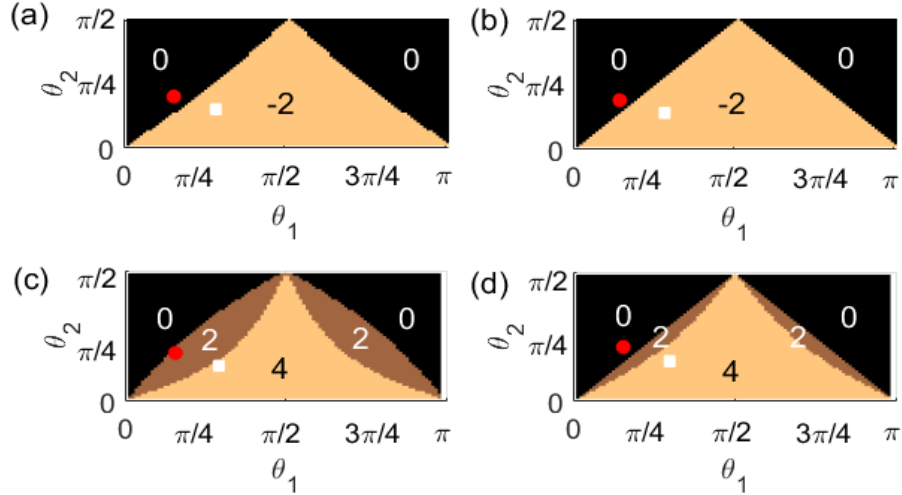

**Fig. 4**. Comparison between the complex phase and the number of edge states. The gain-loss strength in (a) and (c) [(b) and (d)] are 0.3 (0.1). (a) and (b), the complex phase in non-Hermitian QWs. The black and orange regions represent the complex phase values 0 and -2, respectively. (c) and (d), the number of edge states. The black, brown and orange regions denote the value 0, 2 and 4, respectively. Red dots (white rectangles) represent rotation angles $\theta_1 = 0.16\pi$ and $\theta_2 = 0.21\pi$ ($\theta_1 = 0.3\pi$ and $\theta_2 = 0.16\pi$).

In Fig. 4(a), the complex phase is 0 (red dot) for the case with $\theta_1 = 0.16\pi$ and $\theta_2 = 0.21\pi$ when the gain-loss strength $r=0.3$, and it becomes -2 (white rectangle) for the case with $\theta_1 = 0.3\pi$ and $\theta_2 = 0.16\pi$. This means the edge state numbers 0 and 4 at the boundaries of QW if the bulk-boundary correspondence exists. However, the corresponding edge state numbers in Fig. 4(c) are 2 and 4 at the boundaries of non-Hermitian QWs, which are the same to the results shown in Fig. 2(a) and Fig. 3(a), respectively. Therefore, the complex phase value does not agree with the edge state number when rotation angles are taken as

$\theta_1 = 0.16\pi$ and $\theta_2 = 0.21\pi$ at $r = 0.3$. The QW belongs to the broken PT-symmetric phase under these parameters. We also compare the complex phase and edge state numbers at larger r (r=0.6 and 0.8), the difference between these two values for the broken PT-symmetric phase always exists (see S5 of Supplementary Material).

For comparison, the QW belongs to unbroken PT-symmetric phase, the edge state number can be predicted correctly from the complex phase. For example, when the gain-loss strength $r = 0.1$, we take rotation angles $\theta_1 = 0.16\pi$ and $\theta_2 = 0.21\pi$ ( $\theta_1 = 0.3\pi$ and $\theta_2 = 0.16\pi$ ), red dots and white rectangles in Fig. 4(b) and Fig. 4(d) locate at the corresponding region. The numbers of edge states are agreement well with the complex phase values. This means that the edge state number can be predicted correctly from the complex phase for the unbroken PT-symmetric phase.

Moreover, calculation results from other theories (Q-Matrix, the vorticity of energy eigenvalues, topological defects and the step-average of mean position of walker) have also shown that they cannot be used to describe the edge states correctly when the system belongs to the broken PT-symmetric phase. Details of these calculations have been provided in S5 of Supplementary Materials. It is concluded that when the system belongs to the broken PT-symmetric phase, the bulk-boundary correspondence with current characterization methods does not exist. A new efficient method needs to be developed for describing broken PT-symmetric topological phases in non-Hermitian QWs.

## 5. Discussions and conclusions

The topological phenomena related to the non-Hermitian QWs have attracted intense attentions in recent years. The observation of edge states at the interface between two unbroken PT-symmetric phases has been presented in Ref. [37]. The topological transition have been revealed in the one-dimensional lattice with the decay appearing at one of two sublattices [42 43]. Such topological transition has also been generalized to Hermitian [44] and non-Hermitian discrete time QWs [45, 46]. To our knowledge, the demonstration of robust edge states in the broken PT-symmetric phase has not been involved. There is a lack of detailed discussion about whether the bulk-boundary correspondence exists in such non-Hermitian QWs.

In this work, we have experimentally observed unconventional edge states in non-Hermitian QWs with broken PT-symmetric phases. The observations are based on our constructed non-Hermitian QW platform with left and right boundaries, where the beams outside the system boundary are blocked subtly at the end of each step. Thus, these edge states are only caused by the broken PT-symmetric phases with strong dissipation, which are robust against to static perturbations and disorder. The phenomenon cannot be explained by all current theories on topological invariant, although the edge states in unbroken PT-symmetric phases have been proved to be satisfied the bulk-boundary correspondence. Our demonstration of non-existence of bulk-boundary correspondence in non-Hermitian systems will stimulate more efforts in discovering new methods to characterize topological phases. Furthermore, although the dissipation of real systems cannot be avoided in nature, our finding of novel robust edge states will inspire us to explore a robust transport channel [47] in complex systems with strong dissipation.

**Acknowledgement**

This work was supported by the National key R & D Program of China under Grant No. 2017YFA0303800 and the National Natural Science Foundation of China (11574031 and 11604014).

See Supplementary materials for supporting content.